\documentclass[pdflatex,sn-mathphys-num,iicol]{sn-jnl}

\usepackage{graphicx}%
\usepackage{multirow}%
\usepackage{amsmath,amssymb,amsfonts}%
\usepackage{amsthm}%
\usepackage{mathrsfs}%
\usepackage[title]{appendix}%
\usepackage{xcolor}%
\usepackage{textcomp}%
\usepackage{manyfoot}%
\usepackage{booktabs}%
\usepackage{algorithm}%
\usepackage{algorithmicx}%
\usepackage{algpseudocode}%
\usepackage{listings}%
\usepackage{lineno}%

\usepackage{natbib}

\begin{document}

\title{Improvement and assessment of the radiopurity of Micromegas readout planes}




\author[1]{\fnm{Juan} \sur{Castel}}

\author[1]{\fnm{Susana} \sur{Cebrián}}\email{scebrian@unizar.es}

\author[1]{\fnm{Theopisti} \sur{Dafni}}

\author[1]{\fnm{David} \sur{Díez-Ibáñez}}

\author[1]{\fnm{Álvaro} \sur{Ezquerro}}

\author[1]{\fnm{Juan Antonio} \sur{García}}

\author[1]{\fnm{Héctor} \sur{Gómez}}\email{hgomez@unizar.es}

\author[1]{\fnm{Igor} \sur{G. Irastorza}}

\author[1]{\fnm{Gloria} \sur{Luzón}}

\author[1]{\fnm{Cristina} \sur{Margalejo}}

\author[1]{\fnm{Héctor} \sur{Mirallas}}

\author[1]{\fnm{Luis} \sur{Obis}}

\author[2]{\fnm{Rui} \sur{de Oliveira}}

\author[1]{\fnm{Alfonso} \sur{Ortiz de Solórzano}}

\author[1]{\fnm{Óscar} \sur{Pérez}}

\author[1]{\fnm{Jorge} \sur{Porrón}}

\author[1]{\fnm{María} \sur{J. Puyuelo}}

\author[1]{\fnm{Ana} \sur{Quintana}}\email{aquintana@unizar.es}

\author[1]{\fnm{María} \sur{Rodríguez}}

\author[1]{\fnm{Laura} \sur{Seguí}}


\affil[1]{\orgdiv{Centro de Astropartículas y Física de Altas Energías (CAPA)}, \orgname{Universidad de Zaragoza}, \orgaddress{\street{Pedro Cerbuna, 12}, \city{Zaragoza}, \postcode{50009},  \country{Spain}}}

\affil[2]{\orgname{European Organization for Nuclear Research (CERN)}, \orgaddress{\city{Geneva 23}, \postcode{1211}, \country{Switzerland}}}

 
\date{}


\abstract{Micromesh Gas Structures (Micromegas) as readout of gaseous Time Projection Chambers (TPCs) are being considered in experiments investigating rare phenomena, like the nuclear double beta decay, solar axion detection and low-mass dark matter interactions, due to their good performance on spatial and energy resolution and operation stability. In addition, as they are potentially made mainly of radiopure materials like copper and kapton, they are appropriate for ultra-low background conditions. After a promising first study of the radiopurity of Micromegas readout planes, here results after dedicated development at CERN obtained from new radioassays, performed at the Canfranc Underground Laboratory combining different techniques, are presented. Activity of the isotopes in the lower parts of the $^{238}$U and $^{232}$Th natural chains has been constrained by analyzing the BiPo sequences using the BiPo-3 detector to be $<$0.064 and $<$0.016~$\mu$Bq/cm$^2$ respectively, while a lowest $^{40}$K content of 0.102$\pm$0.030~$\mu$Bq/cm$^2$ has been determined by gamma spectroscopy using a HPGe detector; the latter value implies a reduction of a factor $\sim$34 with respect to the $^{40}$K activity quantified in the first analyzed sample. These results confirm the suitability of the use of Micromegas as extremely radiopure readouts for rare event searches.}


\maketitle

\section{Introduction}

In modern Time Projection Chambers (TPCs), MicroPattern Gas Detectors (MPGD) have replaced MultiWire Proportional Chambers (MWPC) readout planes for amplification and measurement of the drifting charge generated in the gas. MPGD use metallic strips or pads, printed on plastic supports with photolithography techniques, improving simplicity, robustness and mechanical precision; different designs have been developed based on different multiplication structures, like the Micromesh Gas Structures (Micromegas) readout planes \cite{MMGiomataris}.

Gas TPCs can provide topological information of events (registered by an appropriately patterned readout) for signal identification and background rejection; this is a very useful capability for the investigation of phenomena expected to occur with extremely low probability, like the neutrinoless double beta decay \cite{GomezCadenas2023}, the direct detection of dark matter particles in the galactic halo \cite{Billard2022} or the interaction of axions \cite{Axreview}. Although MWPC TPCs were not best suited for large detection volume and very long exposure operation, limitations have been greatly overcome with MPGDs and advances in electronics. In particular, Micromegas have shown improved energy resolution, homogeneity, stability of operation and possibility of scaling-up, which are key elements for applying gas TPCs for searches requiring ultralow backgrounds \cite{Cebrian:2010nw,Irastorza:2015dcb,Irastorza:2015geo}. 


Since their conception \cite{MMGiomataris}, Micromegas readouts have been used in many different applications in nuclear, particle and astroparticle physics; as an example, the ATLAS Muon Spectrometer has 1200 m$^2$ of Micromegas planes. In the context of rare event searches requiring low background techniques, they were used for the first time in the CERN Axion Solar Telescope (CAST) at CERN \cite{Anastassopoulos:2017ftl,PhysRevLett.133.221005} and are being considered for the future International Axion Observatory (IAXO) \cite{Abeln2021BabyIAXO}. Micromegas are being considered also in dark matter direct detection experiments like TREX-DM (TPCs for Rare Event eXperiments-Dark Matter) using a High Pressure (HP) TPC \cite{Irastorza:2015geo,Castel:2018gcp,Castel2024TREXDM} and MIMAC (MIcro-tpc MAtrix of Chambers), focused on the signal directionality detection \cite{Guillaudin2025}. The use of Micromegas in the investigation of double beta decay has been also proposed \cite{Cebrian:2010nw,Irastorza:2015dcb} and applied by the PANDAX-III experiment with a HP Xe TPC to study the decay of $^{136}$Xe \cite{pandaxiii}.

For any rare event phenomenon, the searched signal is expected to appear entangled with background events having different origins. Different strategies are followed to suppress the background as much as possible \cite{heusser,formaggio}: use of active and passive shieldings, operation in underground sites to reduce cosmic radiation, implementation of event discrimination techniques and careful selection of materials and components having the lowest possible radioactive traces. The latter implies the assessment of very low levels of material radioactivity by means of different techniques like Neutron Activation Analysis (NAA), Inductively Coupled Plasma Mass Spectrometry (ICPMS), Glow Discharge Mass Spectrometry (GDMS) and gamma spectroscopy performed underground with ultralow background germanium detectors (see recent examples in \cite{scrmaj,screxo,scrnext,scrnexe,Akerib2020LZcleanliness,Aprile2022XENONnTclean});
even very specific techniques have been developed, like the quantification of the so-called BiPo events in the natural radioactive chains \cite{BiPo_detector}. In the exhaustive control of radioactive impurities, bulk and surface contamination are considered and not only primordial or anthropogenic but also cosmogenic activity \cite{cebriancosmogenic,universecosmogenic} must be taken into account.

Therefore, the study of the radiopurity of Micromegas readouts and that of closely-related components to be used in rare event experiments has been addressed in previous works \cite{aznar2013,Iguaz:2015myh}; in particular, the very first screening of different Micromegas samples \cite{Cebrian:2010ta} confirmed that they were much suited for low background applications and that it could be possible to improve radiopurity after dedicated development. Then, the goal of this work has been to implement different changes in the fabrication techniques of Micromegas planes and assess the activity of the relevant radioisotopes using several complementary techniques carried out at the Canfranc Underground Laboratory (LSC, ``Laboratorio Subterráneo de Canfranc''). The structure of the paper is as follows. Section~\ref{fab} presents the process for the fabrication and preparation of all the samples considered. In Section~\ref{act}, the techniques applied to carry out the radioassays are described before showing in detail the measurements performed and the results obtained. The conclusions are gathered in Section~\ref{con}.


\section{Fabrication and sample preparation}
\label{fab}

The Micromegas readouts consist of a metallic micromesh suspended over an anode plane (usually segmented into strips or pixels) by means of insulator pillars, defining an amplification gap of the order of 50 to 150~$\mu$m. In a TPC, the Micromegas planes (instead of wires) play the role of the electron amplifying and sensing electrodes. Electrons drifting towards the readout, go through the micromesh holes and trigger an avalanche inside the gap, inducing detectable signals both in the anode and in the mesh. Low intrinsic gain fluctuations and a relatively low dependence of the gain on geometrical (the gap size) or environmental factors (like the temperature or pressure of the gas) are the advantages of Micromegas with respect to other MPGDs. The manufacturing technology of the Micromegas planes has evolved significantly throughout the years, being facilitated by the development of all-in-one readouts. The bulk Micromegas \cite{MMbulk} uses a photo resistive film to integrate the mesh and anode while microbulk Micromegas \cite{MMmicrobulk} are constructed out of a (typically 50 $\mu$m-thick) kapton foil, doubly-clad with copper, with anode and micromesh patterns engraved; the amplification gap is created by chemically removing part of the kapton layer below the mesh holes. Microbulk Micromegas planes show great geometrical homogeneity, which results in an improved stability and energy resolution. The spatial resolution depends on the granularity of the anode. Additionally, the materials they are made of (kapton and copper) have typically very low levels of intrinsic radioactivity. For all these reasons, microbulk Micromegas are being considered in rare event searches and the study presented here is focused on this type of Micromegas. Although it must be noted that microbulk Micromegas are less robust than the bulk ones and that the maximum size of single readouts is limited.

The radiopurity of Micromegas readout planes was firstly studied in depth in \cite{Cebrian:2010ta}. Samples representative of the raw materials as well as manufactured readouts were analyzed by germanium spectroscopy in the Canfranc Underground Laboratory. On the one hand, two samples were part of fully functional Micromegas detectors: a microbulk readout plane (formerly used in the CAST experiment) and a kapton Micromegas anode structure without mesh. On the other hand, two more samples were just raw foils used in the fabrication of microbulk readouts, consisting of kapton metalized with copper on one or both sides. The raw materials were confirmed to be very radiopure, bounding their contamination to less than tens of $\mu$Bq/cm$^2$ for the natural $^{238}$U and $^{232}$Th chains and for $^{40}$K. Similar limits or values just at the limit of the sensitivity of the measurement were obtained for the treated foils. Despite their relevance, these bounds were still relatively modest when expressed in volumetric terms due to the small mass of the samples. Therefore, more sensitive screening techniques using the BiPo-3 detector and more massive samples were considered for the study of new Micromegas readouts prepared in controlled conditions applying different procedures.

Some of these samples already screened were prepared for analysis at the BiPo-3 detector. One was the Cu-kapton-Cu foil, consisting of four disks (diameter 11~cm) with a total mass of 4.7~g. Copper is 5~$\mu$m thick and kapton 50~$\mu$m thick. The other one was the CAST microbulk Micromegas detector (diameter 11~cm too) with a mass of 2.8~g; it includes several layers of copper, kapton and epoxy. Both samples were protected by polyethylene film for installation in BiPo-3.


First, faulty Micromegas circuits produced at CERN were considered to perform specific analyses both at BiPo-3 and germanium detectors accumulating a larger quantity of material than in the previous samples (sample MM\#0). This sample had not the holes, which are produced using a potassium compound.

Then, a first massive sample (sample MM\#1) was prepared at CERN, consisting of failed GEM glued on kapton, as it is done for microbulk Micromegas. The existence of holes was confirmed by inspection using a microscope. The available sheets were cut into 78 pieces with a surface of 13.5$\times$11.75~cm$^2$ each to fit into the marinelli container which is placed on top of the germanium detectors used for screening.  After a first screening at LSC of the prepared sample, two cleaning processes were successively carried out at CERN consisting of water baths: in the first one, the foils were dipped one week in tap water followed by a long rinse with deionized (DI) water; in the second one, the baths were performed only with DI water, heated to 60$^o$C and changed each day for one week.

Following the results obtained after the processing of sample MM\#1, a new massive sample (sample MM\#2) of microbulk-equivalent Micromegas was prepared at CERN. It consisted of 70 pieces with a surface of 10$\times$10~cm$^2$ each. The following treatments were applied during production: two baths were used containing potassium in the GEM production, in the kapton etching step and in the cleaning step. On the cleaning step, potassium permanganate was used followed by a bath which neutralise this permanganate. The neutralisation time was set to have low leakage current in the GEM. After the first screening at LSC, a cleaning process was carried out at CERN using DI water (avoiding tap water).

Due to the limit of sensitivity achievable on sample MM\#2, a third, very massive sample (sample MM\#3) of microbulk-equivalent Micromegas was prepared at CERN. The available large foils were cut into strips to be placed inside the marinelli container. 

Finally, a study of the radiopurity of a witness sample of the microbulk Micromegas produced at CERN for the TREX-DM experiment \cite{Castel2024TREXDM} was performed (sample MM\#4). It consisted of 75 pieces with a surface of 10$\times$11~cm$^2$ each. The first screening was performed on the sample as received, with a mandatory cleaning with potassium permanganate applied. A second measurement was carried out after cleaning with demineralized water for one week at CERN. Following a second cleaning with potassium permanganate intended to improve Micromegas performance, a third last analysis was made. 

In addition to the Micromegas readouts produced under different conditions,
a number of other samples involved in various Micromegas fabrication processes were also assayed using the same techniques, germanium spectroscopy and the BiPo-3 detector. The screened samples were a vacrel film used in the construction of bulk Micromegas, a sheet of CVD (Chemical Vapor Deposition) diamond‐coated polyimide having big resistivity and then of interest for the development of resistive microbulk Micromegas (consisting of a 50‐$\mu$m‐thick kapton layer coated with 0.1‐$\mu$m‐thick carbon), a kapton-epoxy foil (AKAFLEX, from Krempel GmbH) used to join several kapton layers in more complex routing designs in the fabrication process of microbulk Micromegas (consisting of a 50‐$\mu$m‐thick kapton layer and 21.4‐$\mu$m‐thick epoxy), and adhesive Isotac 3M VHB used in Micromegas readouts.

\begin{figure*}
\centering
\includegraphics[width=0.7\textwidth]{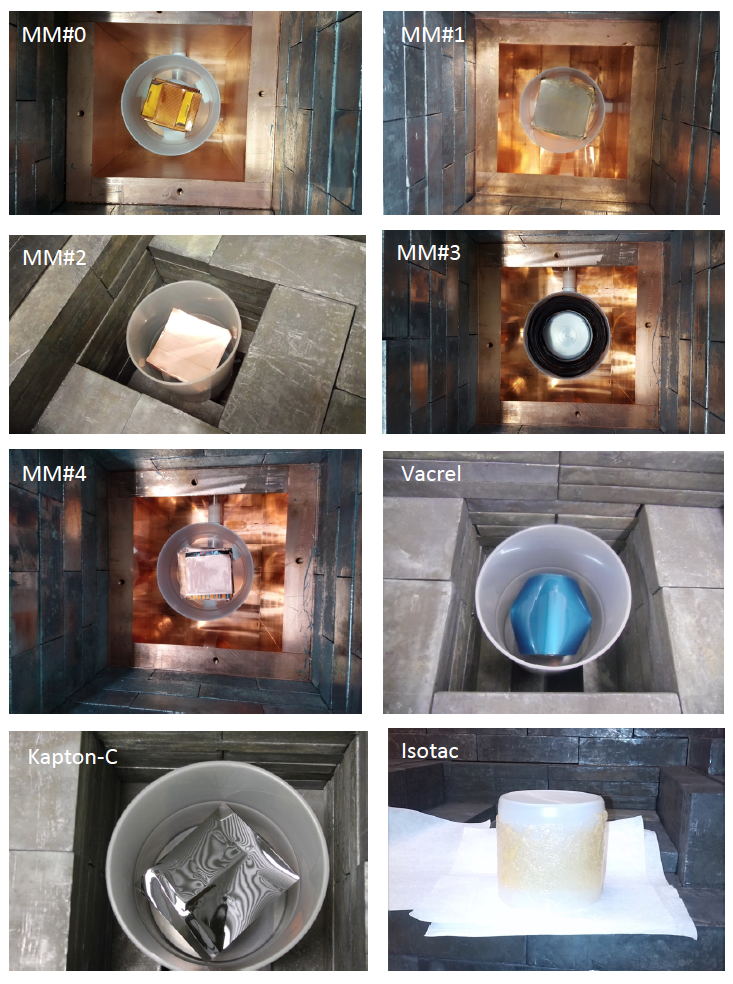}
\caption{\centering Pictures of samples screened with germanium detectors at LSC described in Table~\ref{tabGesamples}.} 
\label{FotosGe}
\end{figure*}

\begin{figure*}
\centering
{
\includegraphics[width=0.7\textwidth]{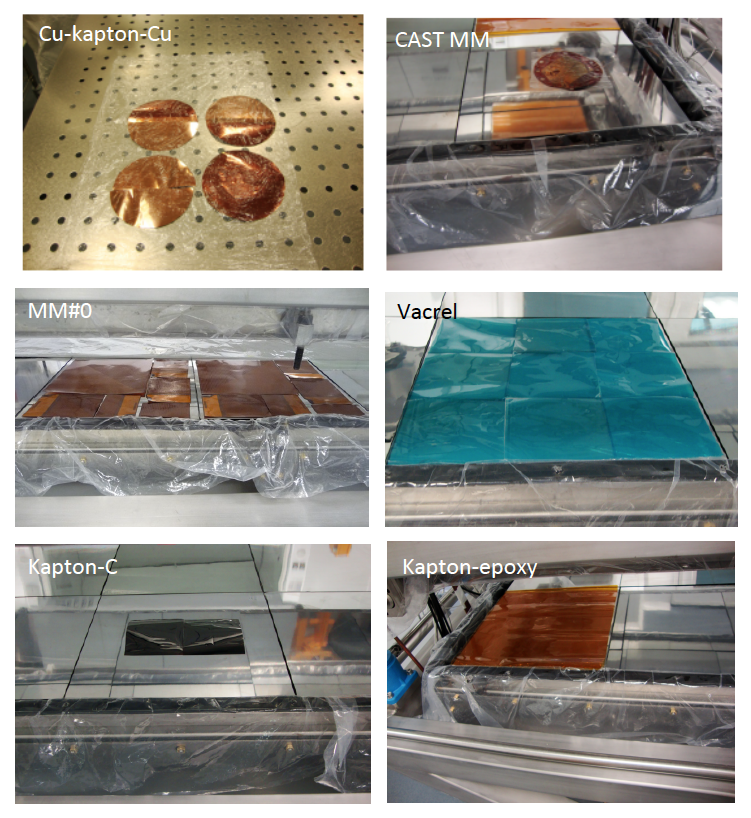}
\caption{\centering Pictures of samples analyzed with the BiPo-3 detector at LSC described in Table~\ref{tabBiPosamples}.} 
}\label{FotosBiPo}
\end{figure*}

The features of all the samples assayed with germanium detectors are given in Table~\ref{tabGesamples} and their pictures are shown in Fig.~\ref{FotosGe}, while those analyzed with the BiPo-3 detector are described in Table~\ref{tabBiPosamples} and depicted in Fig.~\ref{FotosBiPo}.

\begin{table*}
\begin{center}
\caption{Features (mass and surface) of the Micromegas (MM) samples prepared at CERN and other related samples and description (detector used and time of measurement) of the germanium radioassays carried out at LSC.}
\label{tabGesamples} 
\begin{tabular}{lcccccl}
\toprule
Sample & Mass & Surface  & Meas. & Detector & Time & Description  \\
& (g) & (cm$^2$) & & & (d) &\\
\midrule
MM\#0 & 73.8 & 1587 & \#0 & GeLatuca & 50.6 & As produced and measured in BiPo-3 \\ \hline
MM\#1 & 165.9 & 12373 & \#0 & GeOroel & 41.6 & As produced \\
&&&  \#1 & GeOroel & 29.0 & After first cleaning (DI, tap water) \\
&&&  \#2 & GeOroel & 30.9 & After second cleaning (DI water) \\ \hline
MM\#2 & 86.2 & 7000 & \#0 & Paquito & 60.6 & As produced \\
&&&  \#1a & Paquito & 76.8 & After first cleaning (DI water) \\
&&&  \#1b & GeOroel & 46.5 & After first cleaning (DI water) \\ \hline
MM\#3 & 637 & 51735 & \#0 & GeAnayet & 31.9 & As produced \\ \hline
MM\#4 & 116.5 & 8250 & \#0 & GeAnayet & 29.7 & As produced \\
&&&  \#1 & GeAnayet & 58.7 & After first cleaning (DI water) \\
&&&  \#2 & GeAnayet & 30.4 & After second cleaning\\ 
&&& & & & (potassium permanganate)\\ \hline
Vacrel & 65.5 & 4800 & & Paquito & 32.3 & Also measured in BiPo-3 \\ \hline
Kapton-C & 2.8 & 189 & & Paquito & 40.3 & Also measured in BiPo-3 \\ \hline
Isotac & 19.4 & 3202 & & Paquito & 44.4  &  \\ 
\botrule
\end{tabular}
\end{center}
\end{table*}

\begin{table*}[h]
\begin{center}
\caption{Features (mass and surface) of the samples analyzed 
the BiPo-3 detector at LSC and description (time of measurement and simulated efficiency) of the radioassays carried out. The simulated efficiency takes into account the same event selection applied to the real data.}
\label{tabBiPosamples} 
\begin{tabular}{lcccccc}
\toprule
Sample & Mass & Surface & Time  & Efficiency  & Time  & Efficiency  \\
&  & & ($^{212}$BiPo) & ($^{212}$BiPo) & ($^{214}$BiPo) & ($^{214}$BiPo) \\
& (g) & (cm$^2$) &  (d) & (\%) &  (d) & (\%)\\
\midrule
Cu-kapton-Cu & 4.7 & 380 & 158 & 7.22 & 134 & 4.12 \\ 
CAST Micromegas & 2.8 & 95 & 79 & 7.42 & 67 & 4.23 \\ 
MM\#0 & 73.8 & 1587 &  253 & 3.89 & 241 & 0.83 \\ 
Vacrel & 13.1\footnotemark & 900 & 166 & 4.14 & 154 & 2.27 \\ 
Kapton-C & 1.34\footnotemark[\value{footnote}] & 189 & 166 & 13.41 & 154 & 7.80  \\ 
Kapton-epoxy & 9.4 & 900 & 79 & 8.79 & 67 & 7.01 \\ 
\botrule
\end{tabular}
\footnotetext{Estimated values, in the absence of direct measurements.}
\end{center}
\end{table*}


\section{Activity measurements}
\label{act}

The radioassay of the prepared samples has allowed to assess the effectiveness of the different procedures applied and to quantify the radiopurity level of the Micromegas readouts. The very low levels of activity expected have imposed the use of different, complementary techniques. After a brief description of these analysis methods, the measurements carried out will be presented and the obtained results discussed.

\subsection{Techniques}

Germanium gamma spectroscopy measurements have been carried out at the LSC. Some of them have been made using a $\sim$1~kg ultra-low background detector of the University of Zaragoza (named Paquito) operated at the hall LAB2500 of the LSC. This detector has been used for radiopurity measurements at Canfranc for many years (details on the features and performance of the detector can be found in \cite{aznar2013,Cebrian:2010ta}). Additionally, some of the $\sim$2.2~kg close-end coaxial HPGe detectors of the UltraLow Background Service of LSC \cite{ulbslsc} have been used for other measurements too. All detectors are operated inside heavy shieldings made of copper and/or low activity lead, enclosed in plastic boxes continuously flushed with boil-off nitrogen to avoid radon intrusion. Activities of different sub-series in the natural chains of $^{238}$U, $^{235}$U and $^{232}$Th as well as of common primordial, cosmogenic or anthropogenic radionuclides like $^{40}$K, $^{60}$Co and $^{137}$Cs are typically evaluated, following the procedure applied in \cite{Baudis_2011}. The detection efficiency is determined by Monte Carlo simulations based on Geant4 (validated with a $^{152}$Eu reference source) for each sample, taking into account composition and geometry; a conservative overall uncertainty on the deduced efficiency is properly propagated to the final results.

The BiPo-3 detector \cite{BiPo_detector} was developed by the SuperNEMO collaboration to measure the extremely low levels of $^{208}$Tl and $^{214}$Bi, produced in the decays of the natural chains of $^{232}$Th and $^{238}$U, in the foils that contain the double beta decay emitter studied in the experiment. Placing the sample foil between two thin layers of scintillators, it is possible to register the BiPo events from the chains by detecting the electron energy deposition in one of the detectors and the delayed alpha signal in the opposite one. The detector consisted of two modules containing 40 optical submodules (positioned in two rows), having a sensitive surface of 3.6~m$^2$. Each capsule had two organic plastic polystyrene-based scintillators (30$\times$30$\times$0.3~cm$^3$ each) coupled to low radioactive 5-inch photomultipliers (PMTs, model Hamamatsu R6594-MOD) through PMMA optical guides to register the energy and time of the energy depositions. MATACQ boards were used to sample the PMT signals. The modules were held in a pure iron structure, placed inside a radon-tight stainless steel tank containing also the low activity lead shielding for a total thickness of 10~cm. The insertion of the samples inside the BiPo-3 detector was performed in the LSC clean room. Operated at the LSC for several years before dismantling, BiPo-3 was able to reach sensitivities down to the few $\mu$Bq/kg level. This method is adequate for any sample in the form of a thin foil (below 100-200~$\mu$m thick, so that some alphas can escape the sample) as is the case of Micromegas readouts; the BiPo analysis has been applied also in this study.


The BiPo-3 data were analyzed here using the REST-for-physics code \cite{REST_2022}, following the procedure established by the SuperNEMO collaboration \cite{BiPo_detector}. This method was previously validated with a calibrated aluminum foil. The event selection criteria are summarized in Table~\ref{BiPo_cuts}. The first cuts impose energy thresholds on the prompt ($\beta$) and delayed ($\alpha$) signals, followed by a coincidence time window. A crosstalk rejection cut excludes induced signals between opposite channels of a capsule. Pulse shape discrimination for PMT noise rejection is enforced through a constraint on the charge over amplitude ratio (Q/A). Finally, data from the first days after each detector opening are discarded to suppress radon-induced backgrounds; the rejection period is determined from the time distribution of events, which is also checked for possible correlations. Contrary to previous presentations of BiPo-3 results \cite{Castel:2018gcp, aznar2013}, a Bayesian statistical approach has been used in this work instead of the Feldman–Cousins method.

\begin{table*}[h]
    \begin{center}
    \caption{Cuts applied for the selection of $^{212}$BiPo and $^{214}$BiPo events.}
    \label{BiPo_cuts}
    \begin{tabular}{lcc}
    \toprule
    & $^{212}$BiPo & $^{214}$BiPo \\ 
    \midrule
    Threshold $\text{E}_{\text{prompt}}$ & 200\,keV  & 200\,keV \\
    Threshold $\text{E}_{\text{delay}}$ & 300\,keV (100$-$700\,keV for MM\#0) & 300\,keV (300$-$600\,keV for MM\#0) \\
    $\Delta t$ & 50 $-$ 1400 ns & 1.5 $-$ 1000 $\mu$s \\
    Crosstalk & 10\,mV & 10\,mV \\
    Q/A & Mean $\pm$ 3$\sigma$ & Mean $\pm$ 3$\sigma$ \\
    First days rejection & 3 days & 15 days \\
    \botrule
    \end{tabular}
    \end{center}
\end{table*}

\subsection{Measurements}

Firstly, profiting from the great capabilities of the BiPo-3 detector operating at LSC, two of the samples considered in the study in \cite{Cebrian:2010ta} (Cu–kapton–Cu foil sample and CAST microbulk Micromegas) were analyzed again using this detector.

A more sensitive measurement was prepared for sample MM\#0, using two capsules of the BiPo-3 detector. 
After the analysis in the BiPo-3 detector, the sample was cut into several pieces and screened using one germanium detector of the UltraLow Background Service of LSC (the one named GeLatuca) mainly to analyze the $^{40}$K activity (measurement \#0).

Another germanium detector of the UltraLow Background Service of LSC (the one named GeOroel) was used for all the analyses of sample MM\#1; three measurements were performed: one just after the preparation of the sample (measurement \#0) and two after each one of the cleaning processes at CERN (measurements \#1 and \#2).

For sample MM\#2, the screening after preparation was performed using the Paquito detector (measurement \#0). After the cleaning with DI water, two new radiopurity measurements were carried out, one using also the Paquito detector (measurement \#1a) and then other one using the GeOroel detector in order to improve the sensitivity (measurement \#1b). 

For sample MM\#3, the screening after preparation was performed using one germanium detector of the UltraLow Background Service of LSC (the one named GeAnayet) (measurement \#0).

This detector was also used for all the measurements for sample MM\#4, as produced (measurement \#0), after the first cleaning with DI water (measurement \#1) and after the second cleaning with potassium permanganate (measurement \#2).

The Paquito detector was used for the samples made of Vacrel, Kapton-Carbon and Isotac adhesive. A part of the Vacrel sample and the full Kapton-Carbon one were also measured with the Bipo-3 detector. The kapton-copper samples was also analyzed with the BiPo-3 detector.

The main features of all the measurements carried out at LSC for this study are presented in Table~\ref{tabGesamples} for germanium detectors and in Table~\ref{tabBiPosamples} for the BiPo-3 detector.

\subsection{Results}

Table~\ref{tabgeres} presents the results obtained from germanium spectroscopy and BiPo analysis for all the samples and measurements. It is worth noting that, in comparison with preliminary values presented for instance in \cite{pandaxiii,Irastorza:2015dcb}, final values shown here from the BiPo-3 detector include the full data exposure collected and all have been obtained with the same analysis based on REST-for-physics.

\begin{sidewaystable*}
\begin{center}
\caption{Activity (in $\mu$Bq/cm$^2$) of common radiosotopes for the Micromegas readouts prepared at CERN and related samples, analyzed at LSC using germanium and BiPo-3 detectors. Values reported for $^{238}$U and $^{232}$Th correspond to the upper part of the chains and those of $^{226}$Ra and $^{228}$Th to the lower parts. For germanium spectroscopy results, reported values include $1\sigma$ uncertainties and upper limits are evaluated at 95\% C.L. Values from BiPo-3 detector come from the deduced activities of $^{214}$Bi and $^{208}$Tl isotopes and limits correspond to 95\% C.L. too.}
\label{tabgeres} 
\begin{tabular}{lcccccccc}
\toprule
Sample, Measurement & $^{238}$U & $^{226}$Ra & $^{232}$Th & $^{228}$Th & $^{235}$U & $^{40}$K  & $^{60}$Co & $^{137}$Cs \\ 
\midrule
Cu-kapton-Cu foil, BiPo-3 && $<$0.038 && $<$0.043 &&& \\ \hline
CAST MM, BiPo-3 && $<$0.39 && $<$0.20 &&& \\ \hline
MM\#0, \#0 & $<$49 & $<$0.70 & $<$1.2 & $<$0.35 & $<$0.22 & $<$2.3 & $<$0.14 & $<$0.13 \\ 
MM\#0, BiPo-3 && $<$0.064 && $<$0.016 &&& \\ \hline
MM\#1, \#0 & $<$5.2 & $<$0.10 & $<$0.22 & $<$0.08 & $<$0.03 & 3.45$\pm$0.40 & $<$0.02 & $<$0.02 \\
MM\#1, \#1 & 7.41$\pm$0.81 & $<$0.21 & 0.19$\pm$0.05 & $<$0.11 & 0.36$\pm$0.04 & 0.84$\pm$0.16 & $<$0.02 & $<$0.03 \\
MM\#1, \#2 & 7.87$\pm$0.89 & $<$0.17 & 0.14$\pm$0.04 & 0.07$\pm$0.02 & 0.36$\pm$0.04 & 0.81$\pm$0.15 & $<$0.03 & $<$0.02 \\ \hline
MM\#2, \#0 & $<$11 & $<$0.28 & $<$0.26 & $<$0.17 & $<$0.11 & 1.80$\pm$0.65 & $<$0.06 & $<$0.08 \\ 
MM\#2, \#1a & $<$11 & $<$0.25 & $<$0.26 & $<$0.15 & $<$0.14 & $<$1.1 & $<$0.06 & $<$0.07 \\
MM\#2, \#1b & $<$4.1 & 0.140$\pm$0.017 & $<$0.19 & $<$0.068 & $<$0.04 & 0.43$\pm$0.13 & $<$0.03 & $<$0.02 \\ \hline
MM\#3, \#0 & $<$2.5 & 0.035$\pm$0.004 & $<$0.067 & $<$0.018 & 0.022$\pm$0.004 & 0.102$\pm$0.030 & $<$0.006 & $<$0.008 \\ \hline
MM\#4, \#0 & $<17$ & $<0.19$ & $<0.58$ & $<0.14$ & $<0.13$ & 1.07$\pm$0.23 & $<0.03$ & $<0.04$ \\
MM\#4, \#1 & $<12$ & $<0.12$ & $<0.49$ & $<0.10$ & $<0.07$ & $<0.63$ & $<0.03$ & $<0.03$ \\
MM\#4, \#2 & $<16$ & $<0.13$ & $<0.41$ & $<0.14$ & $<0.10$ & $<1.2$ & $<0.03$ & $<0.05$ \\ \hline
Vacrel & $<19$ & $<0.61$ & $<0.63$ & $<0.72$ & $<0.19$ & 4.6$\pm$1.9 & $<0.10$ & $<0.14$ \\
Vacrel, BiPo-3 && $<$0.037 && $<$0.024 &&& \\ \hline
Kapton-C & $<350$ & $<13$ & $<14$ &  $<11$ & $<6$ & $<51$ & $<2$ & $<3$ \\
Kapton-C, BiPo-3 && $<$0.070 && $<$0.035 &&& \\ \hline
Isotac & $<18$ & $<0.45$ & $<0.43$ & $<0.22$ & $<0.18$ & $<2.3$ & $<0.10$ & $<0.14$ \\ \hline
Kapton-epoxy, BiPo-3 && $<$0.030 && $<$0.024 &&& \\ 
\botrule
\end{tabular}
\end{center}
\end{sidewaystable*}


The BiPo-3 analysis of the CAST MM sample gave upper limits on the lower part of both $^{238}$U and $^{232}$Th chains at the level of tenths of $\mu$Bq/cm$^2$, but the intrinsic activity of base raw materials seems to be even better, as the upper limits derived for the Cu-kapton-Cu foil also from BiPo-3 analysis were one order of magnitude lower.

In the germanium analysis of sample MM\#0, upper limits were set for all the common radioisotopes. Those for $^{232}$Th and $^{238}$U were, as expected, higher than those obtained from BiPo-3 by more than one order of magnitude. For $^{40}$K, a limit of $<$2.3~$\mu$Bq/cm$^2$ was derived. In particular, the BiPo-3 analysis of this sample yielded upper limits on the lower part of both $^{238}$U and $^{232}$Th chains of $<$64 and $<$16 $n$Bq/cm$^2$, respectively. 

In the first screening (measurement \#0) of sample MM\#1, upper limits were deduced for all the common radioisotopes except for $^{40}$K; the quantified activity (3.45$\pm$0.40)~$\mu$Bq/cm$^2$ (corresponding to a specific activity of (258$\pm$30)~mBq/kg), is larger than the upper limit set for sample MM\#0, having no holes according to the inspection using a microscope; this seems to confirm that the potassium content is related to the production of holes in the readout.

After the first cleaning procedure (measurement \#1) of sample MM\#1, the following results can be highlighted:
\begin{itemize}
\item Important activities for the upper parts of the radioactive chains were measured. The ratio of activities of $^{238}$U and $^{235}$U is 20.6, in very good agreement with the expectation for natural uranium. The origin of the relevant uranium contamination found, at the level of (564$\pm$62)~mBq/kg of $^{238}$U or (45.8$\pm$5.0)~ppb of U assuming uniform distribution in the sample volume, is not clear. To confirm that it was actually related to the sample, several checks were done: the detector background was checked to not have changed, the marinelli container was separately screened, and compatible results were obtained in a fast screening using other germanium detector. The most plausible hypothesis is that the uranium contamination is related to the water baths made; typical levels of uranium in drinking water are $\sim$10~ppb, being possible a large dispersion around this value. No indication of a possible $^{210}$Pb contamination was found as no excess over background at the 46.5~keV peak was observed, neither before nor after the cleaning procedures.
\item Presence of $^7$Be (T$_{1/2}$=53.22 days, with electron capture decay to ground or excited states) was identified too thanks to the 477.6~keV gamma emission. 
\item Regarding the $^{40}$K content, after the cleaning procedure the activity was reduced by about a factor 4, down to (0.84$\pm$0.16)~$\mu$Bq/cm$^2$.
\end{itemize}

After the second cleaning process (avoiding the use of tap water, measurement \#2), no relevant change in the derived activities was observed, which points to a null effect of the procedure for both the K and U content.

For the sample MM\#2 prepared at CERN, only the $^{40}$K content was quantified at (1.80$\pm$0.65)~$\mu$Bq/cm$^2$ (corresponding to a specific activity of (146$\pm$53)~mBq/kg) and upper limits were set for all the other common radioisotopes in the first screening made (measurement \#0). 
After the cleaning with DI water (measurement \#1), the following facts can be remarked:
\begin{itemize}
\item A small activity of $^{226}$Ra was quantified, higher than the upper limit set for microbulk Micromegas using the BiPo-3 detector; the origin of this activity is unclear. 
\item As happened after cleaning of sample \#1, presence of $^7$Be was identified.
\item No $^{40}$K activity was quantified using Paquito detector, setting an upper limit at the detection limit of the detector used of $<$1.1~$\mu$Bq/cm$^2$ (corresponding to $<$86 mBq/kg). But the second screening using GeOroel detector allowed to quantify this  activity at a level of (0.43$\pm$0.13)~$\mu$Bq/cm$^2$; this means a reduction of about a factor 4 comparing with the value obtained in the first screening of the sample. This reduction due to the applied treatment is fully compatible with the one obtained in sample \#1 after the first cleaning (although the starting value in sample \#2 was lower than in sample MM\#1). In this measurement using the GeOroel, the detection limit for $^{40}$K was 29 mBq/kg corresponding to 0.35~$\mu$Bq/cm$^2$. Consequently, it was concluded that if a further reduction of $^{40}$K activity was achieved by an additional treatment on that sample, it would be very difficult to quantify the potassium content; this made necessary the preparation of a third, more massive sample.
\end{itemize}

From the screening of sample MM\#3, the following results must be pointed out:
\begin{itemize}
\item Regarding the natural radioactive chains, the activity of $^{226}$Ra was quantified, but at a level even lower than the upper limit set for microbulk Micromegas in sample MM\#0 using the BiPo-3 detector for $^{214}$Bi.
The limit set for the lower part of the $^{232}$Th chain was comparable while slightly higher than the BiPo-3 limit. The activity of $^{235}$U was also quantified, but with a value more than one order of magnitude smaller than the activity measured for sample MM\#1 after cleaning.
\item As for the other samples \#1 and \#2, presence of $^7$Be was identified. 
\item $^{40}$K activity was quantified as (0.102$\pm$0.030)~$\mu$Bq/cm$^2$, corresponding to (8.3$\pm$2.4)~mBq/kg. This is the lowest value measured for any of the Micromegas samples analyzed. Compared to the value obtained in the first screening of sample MM\#1, a reduction of a factor 34 is observed.
\end{itemize}

In the first screening (measurement \#0) for sample MM\#4 no radioisotope, apart from $^{40}$K, was quantified and upper limits were set for their activities. The activity of $^{40}$K was (1.07$\pm$ 0.23)~$\mu$Bq/cm$^2$ (corresponding to (76$\pm$16)~mBq/kg), slightly lower but compatible within uncertainties with the initial measured value for sample MM\#2. After the first cleaning (measurement \#1), $^{40}$K activity was reduced, as expected from the previous MM samples; but the activity value could not be quantified and only an upper limit of $<$0.63~$\mu$Bq/cm$^2$ (corresponding to $<$45~mBq/kg) was set. According to measurement \#2, the second cleaning did not produce a measurable increase of the $^{40}$K activity despite the use of potassium permangante and an upper limit comparable to the activity quantified in the initial measurement was derived.

Concerning the other related samples, for those analyzed both with germanium and BiPo-3 detectors (vacrel and kapton-C foils), only the activity of $^{40}$K was quantified in vacrel and upper limits were set for all the other radioisotopes from germanium spectroscopy; for the lower part of $^{238}$U and $^{232}$Th chains, the limits set from BiPo-3 were significantly more restricted, by more than one order of magnitude in all the cases. For the adhesives, only upper limits were derived from the germanium measurement in the case of Isotac and from BiPo-3 detector for the kapton-epoxy foil.

It is worth noting that an extensive screening campaign was additionally performed addressing other components or materials that are typically associated with the implementation of Micromegas readouts in TPC set-ups (related to gas vessel, field cage, radiation shielding or electronic acquisition system): samples from different suppliers of lead, used for shielding, and copper, used for mechanical and electric components; many different types of electronic connectors, resistors, HV and signal cables and PCB substrates; teflon components and adhesives (see for instance \cite{Irastorza:2015dcb,Castel:2018gcp,aznar2013,Iguaz:2015myh}). 

\section{Conclusions}
\label{con}

The assessment of improved radiopurity of Micromegas planes, used as readout of gaseous Time Projection Chambers (TPCs) in experiments investigating rare phenomena, has been presented. Radioassays have been performed at the Canfranc Underground Laboratory combining different techniques, gamma-ray spectroscopy, performed with several ultra-low background germanium detectors, and the analysis of the BiPo events, produced in the natural chains of $^{238}$U and $^{232}$Th using the BiPo-3 detector, specifically conceived to quantify them. The measurements have been made in parallel to a dedicated development at CERN intended to reduce in particular the $^{40}$K activity, related to the implementation of production procedures using potassium compounds.

The comparison of results obtained for samples analyzed both with germanium and BiPo-3 detectors confirms the better sensitivity of the latter for the lower part of $^{238}$U and $^{232}$Th, having obtained in all cases limits lower by more than one order of magnitude from the BiPo analysis. Most of the upper limits derived in this work from the BiPo-3 detector are of a few tens of nBq/cm$^2$. Nonetheless, the germanium screening has been essential too, not only to study $^{40}$K activity but also to have information on the upper part of the chains (which for instance allowed to identify the contamination produced when using tap water). 

Radioactivity measured for different raw materials used in Micromegas production has been checked to be very good. For the produced microbulk Micromegas, the final lowest measured $^{40}$K activity is (0.102$\pm$0.030)~$\mu$Bq/cm$^2$, a factor 34 lower than the value quantified at the beginning of the development. The quantification of this small activity has been possible thanks to the production of a very massive sample and the use of a sensitive germanium detector. For the lower part of $^{238}$U and $^{232}$Th, the BiPo analysis has yielded upper limits of 64 and 16~nBq/cm$^2$, respectively.


\backmatter
\bmhead{Acknowledgements}

Some of the results presented in this work were made possible thanks to the radiopurity measurements of several samples performed with the BiPo-3 detector. The authors would like to thank the BiPo-3 team within the SuperNEMO collaboration for carrying out these measurements. 
The authors acknowledge support from the European Research Council (ERC) under the European Union’s Horizon 2020 research and innovation programme (ERC-2017-AdG IAXO+, grant agreement No. 788781), from the Agencia Estatal de Investigación (AEI) under the grant agreements PID2019-108122GB-C31 and PID2022-137268NB-C51 funded by MCIN/AEI/10.13039/501100011033/FEDER, as well as funds from “European Union  Next Generation EU/RTR” (Planes complementarios, Programa de Astrofísica y Física de Altas Energías) co-funded by Gobierno de Aragón.
The authors would like to acknowledge the use of the Servicio
General de Apoyo a la Investigación-SAI, Universidad
de Zaragoza, and technical support from LSC and
GIFNA staff.

\bibliography{refs.bib}

\end{document}